\documentclass[%
reprint,
 aps,
 pre,
floatfix,
]{revtex4-1}
\usepackage{graphicx}
\usepackage{textcomp}
\usepackage{gensymb}
\usepackage{dcolumn}
\usepackage{bm}
\usepackage{oPlotSymbl} 

\usepackage{amsmath}
\usepackage{amsthm}
\usepackage{amsfonts}
\usepackage{lipsum}

\begin{document}

\preprint{APS/123-QED}

\title{Predicting transition from selective withdrawal to entrainment in two fluid stratified systems}

\author{Sabbir Hassan}
\author{C. Dalton McKeon}
\author{Darryl James}
 \email{darryl.james@ttu.edu}
\affiliation{%
 Department of Mechanical Engineering, Texas Tech University, Lubbock, Texas 79409\\
}%

\date{\today}

\begin{abstract}
Selective withdrawal is a desired phenomenon in transferring oil from large caverns in US Strategic petroleum reserve (SPR), because entrainment of oil at the time during withdrawal poses a risk of contaminating the environment. Motivated to understand selective withdrawal in an SPR-like orientation, experiments were performed in order to investigate the critical submergence depth as a function of critical flow rate. For the experiments, a tube was positioned through a liquid-liquid interface that draws the lower liquid upwards, avoiding entrainment of the upper fluid. Analysis of the normal stress balance across the interface produced a Weber number, utilizing dynamic pressure scaling, that predicted the transition to entrainment. Additionally, an inviscid flow analysis was performed assuming an ellipsoidal control volume surface for the iso-velocity profile that produced a linear relationship between the Weber number and the scaled critical submergence depth. This analytical model was validated using the experimental data resulting in a robust model for predicting transition from selective withdrawal to entrainment.
\end{abstract}


\maketitle


\section{\label{sec:intro}Introduction\protect\\ }

This paper presents experimental data and discusses results using scale analysis and a simplified inviscid analysis that yields a new relationship between two nondimensional terms that predicts the transition from selective withdrawal to viscous entrainment in a two-liquid, immiscible system.

\begin{figure}
\includegraphics[width=86mm]{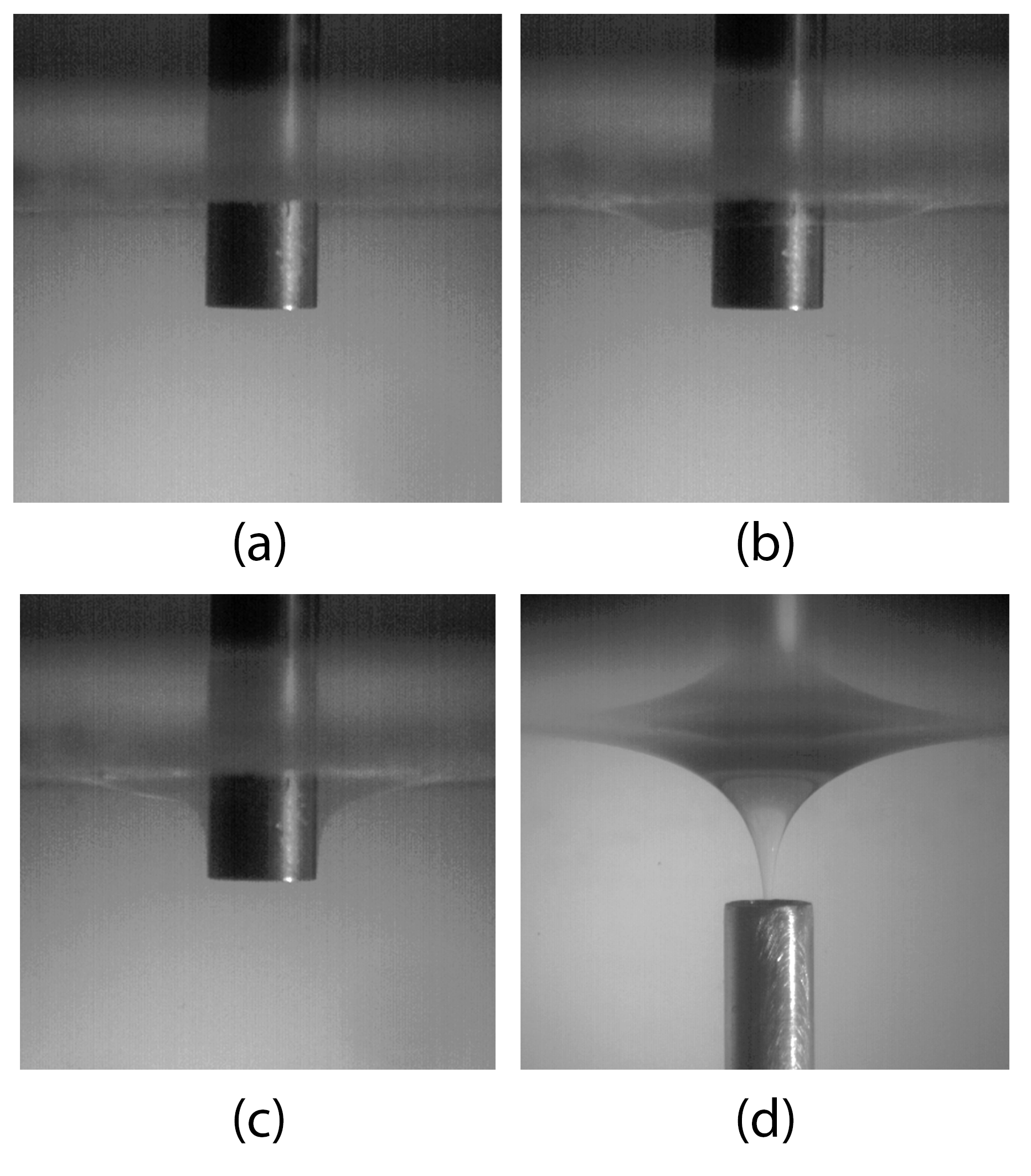}
\caption{\label{fig:1} (a) Primary orientation in an initial state. (b) Primary orientation in a steady, intermediate sub-critical flow rate state. (c) Primary orientation investigated right just after the inception of entrainment. (d) Secondary orientation investigated right just after the inception of entrainment.}
\end{figure}

In the US strategic petroleum reserve (SPR), crude oil is stored in underground salt caverns, which have diameters of approximately 60 m and depths of approximately 600 m \cite{hartenberger2011}. In order to add oil to a cavern, brine is pumped out using a pipe that has been lowered through the interface into the brine. At some critical combination of flow rate and depth below the undeformed interface and the inlet to the pipe, oil is also withdrawn during pumping. This marks the transition from selective withdrawal to entrainment. Selective withdrawal, i.e. removal of only the brine, is desirable during this process, because the removed brine is stored as surface water and any oil in it acts as a pollutant. Motivated to understand selective withdrawal in an SPR-like orientation for which we found no published studies, experiments were performed in order to investigate the critical submergence depth as a function of critical flow rate. For the experiments, a tube was positioned through a liquid-liquid interface that draws the lower liquid upwards, avoiding entrainment of the upper fluid,  i.e. the SPR-like orientation shown in Fig.~\ref{fig:1}(a-c). 

In this paper, we present two approaches to obtain predictive results for the transition from selective withdrawal to entrainment. The first applies dimensional analysis on the interfacial stress balance, while the second uses Bernoulli's Principle or Euler's equation for inviscid fluid flow. The paper is organized as follows: background on selective withdrawal phenomena is provided in Section 2; a brief description of the experiments conducted is given in Section 3; the analytical methods and the necessary fit equations are explained in Section 4; and finally the paper concludes with discussion of the results and the conclusion in Section 5 and Section 6, respectively.

\section{\label{sec:bckgrnd}Background\protect\\ }

Early research conducted on selective withdrawal of two-layer stratified liquids was an experimental study by Lubin and Springer \cite{lubin}. The critical height between the floor of a two-liquid tank and the liquid-liquid interface above was predicted as a function of the sink flow rate. Assuming surface tension to be negligible and using the Bernoulli's principle, a Froude number based on the sink tube diameter was found to be the characteristic nondimensional parameter representing the flow, and a linear relationship between the nondimensional flow rate and the scaled critical depth for their experimental data collapsed their data well.

A more recent experiment was performed by Cohen and Nagel \cite{cohen2002} for a liquid-liquid system in which the upper liquid was withdrawn at a fixed rate (similar to secondary orientation but with the withdrawal tube above the interface). For the given system, the authors found the height of the tube above the interface at the moment of entrainment to be proportional to the volumetric flow rate raised to approximately the 0.3 power. The authors were also able to find a scaling relationship that collapsed all hump profiles to a single profile for low Reynolds number flow. A capillary number based on the capillary length was found to be the representative nondimensional number for the flow. Cohen \cite{cohen2004scaling} extended the research by including several pairs of fluids resulting in different density and viscosity ratios. Based on the low Reynolds number assumption, a dimensional analysis led to a relationship between a scaled flow rate and a scaled distance to the undisturbed interface for prediction of the transition to entrainment. The fit equation had the form of a power law with the exponent varying from 0.30 to 0.45, depending on the fluid combinations used in the experiment. Case and Nagel \cite{case2007} reported a similar relationship in their work while trying to analyze and collapse the spout profiles after entrainment. According to the authors, the spouts have two asymptotic regimes, based on the viscosity ratio, which was matched to each other by the flow dynamics. At large radius the interface is constrained by gravity to be horizontal and at large heights the spout interface is constrained by the flows in the nozzle to be vertical. The nondimensional flow rate based on the capillary number was identified as the representative parameter that collapsed the spout states. 

Blanchette and Zhang \cite{blanchette2009} developed a force balance model to evaluate the system studied experimentally in Cohen paper \cite{cohen2004scaling}.  According to the authors, the transition to entrainment was dependent on a global force balance on the interface, when the upwards force exerted by viscosity because of the withdrawal flow overcomes the downwards force of surface tension. The results of the simulation matched the transition trends found in \cite{cohen2004scaling} closely. It was concluded that the interfacial tension was dominated by the weakly deflected portion of the hump far away from the tip.

In addition to experimental work, the entrainment problem has been investigated numerically. Lister \cite{lister1989} performed a numerical simulation of a two liquid system of equal viscosity in which a point sink was located a distance above an undeformed interface. Assuming Stokes flow and equal viscosity in both layers, the flow field was solved as a function of capillary number, the sink strength, and a viscous velocity scale. A linear trend in the log-log plot was identified between the nondimensional capillary length and sink strength.

Farrow and Hocking \cite{farrow_hocking_2006} used a two dimensional finite-difference approach to simulate selective withdrawal of water for high Reynolds number, inviscid, irrotational flow in order to investigate the scatter in the critical drawdown Froude number, i.e. transition from selective withdrawal. Their results indicated that interfacial waves could affect the critical drawdown Froude number and were responsible for the experimental scatter observed. The critical drawdown Froude number based on the depth of the withdrawn fluid layer, for conditions in which interfacial waves was not a significant factor, is approximately 1. Later, Hocking et al. \cite{hocking_surfacetension_2016} included the surface tension effect due to the curvature of the interface and concluded that the Froude number, based on the sink depth from the interface, represented the complete flow phenomenon and the surface tension adds a resistance to the withdrawal force. However, no predictive relationship between the submergence depth and the critical flow rate was found.

\begin{figure*}
\includegraphics[width=160 mm]{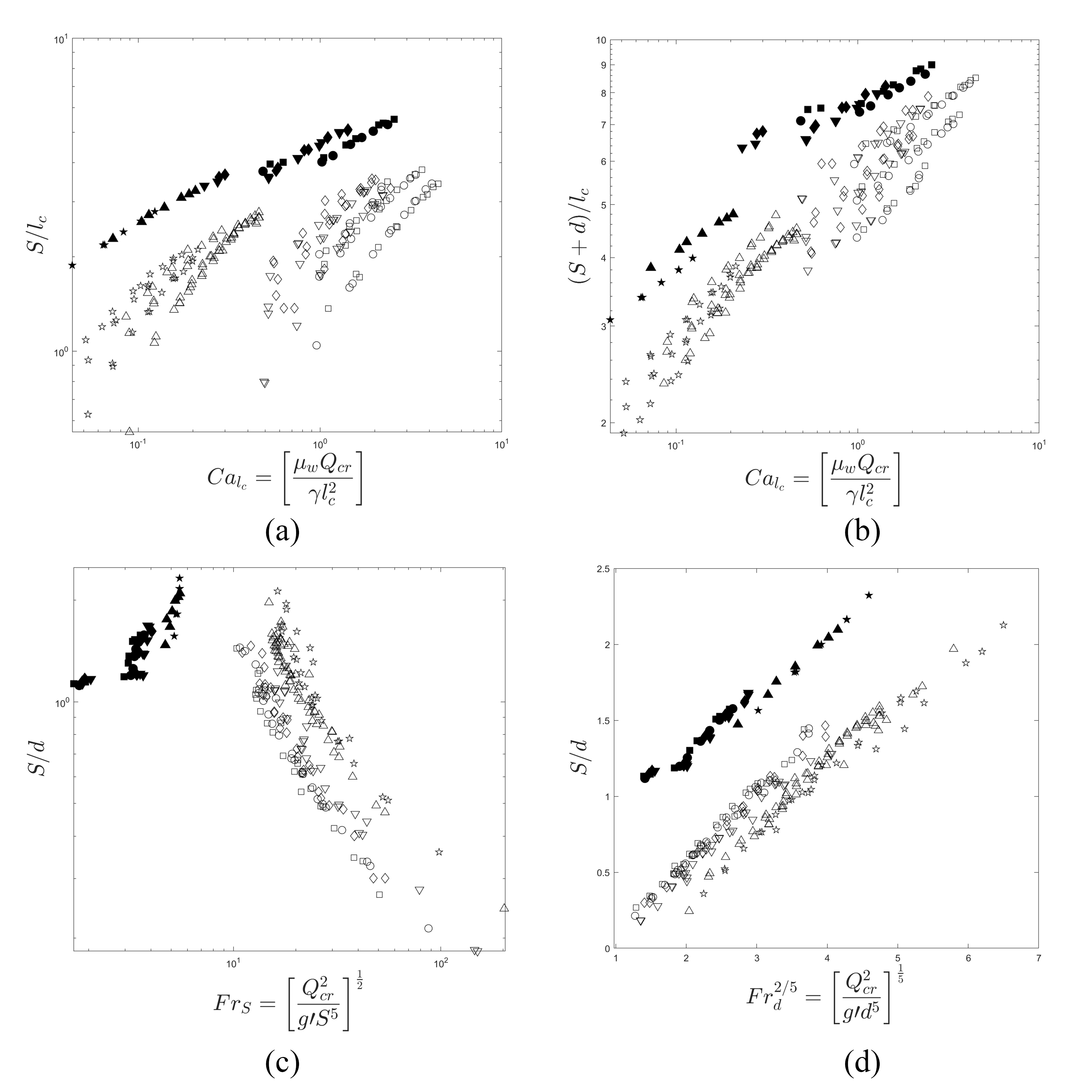}
 \caption{\label{appimage}Plots showing nondimensional submergence depth as a function of various nondimensional parameters. The secondary orientation data are presented in ``filled" symbols, whereas the primary orientation data were presented in ''hollow" symbols. The symbols are defined in Table~\ref{tab:table1}. (a) nondimensional submergence depth as a function of capillary number as in \cite{cohen2004scaling}. (b) nondimensional submergence depth as a function of capillary number as in \cite{blanchette2009}. (c) nondimensional submergence depth as a function of capillary number as in \cite{farrow_hocking_2006}. (d) nondimensional submergence depth as a function of capillary number as in \cite{lubin}.}
\end{figure*}

\begin{figure}
\includegraphics[width=86mm]{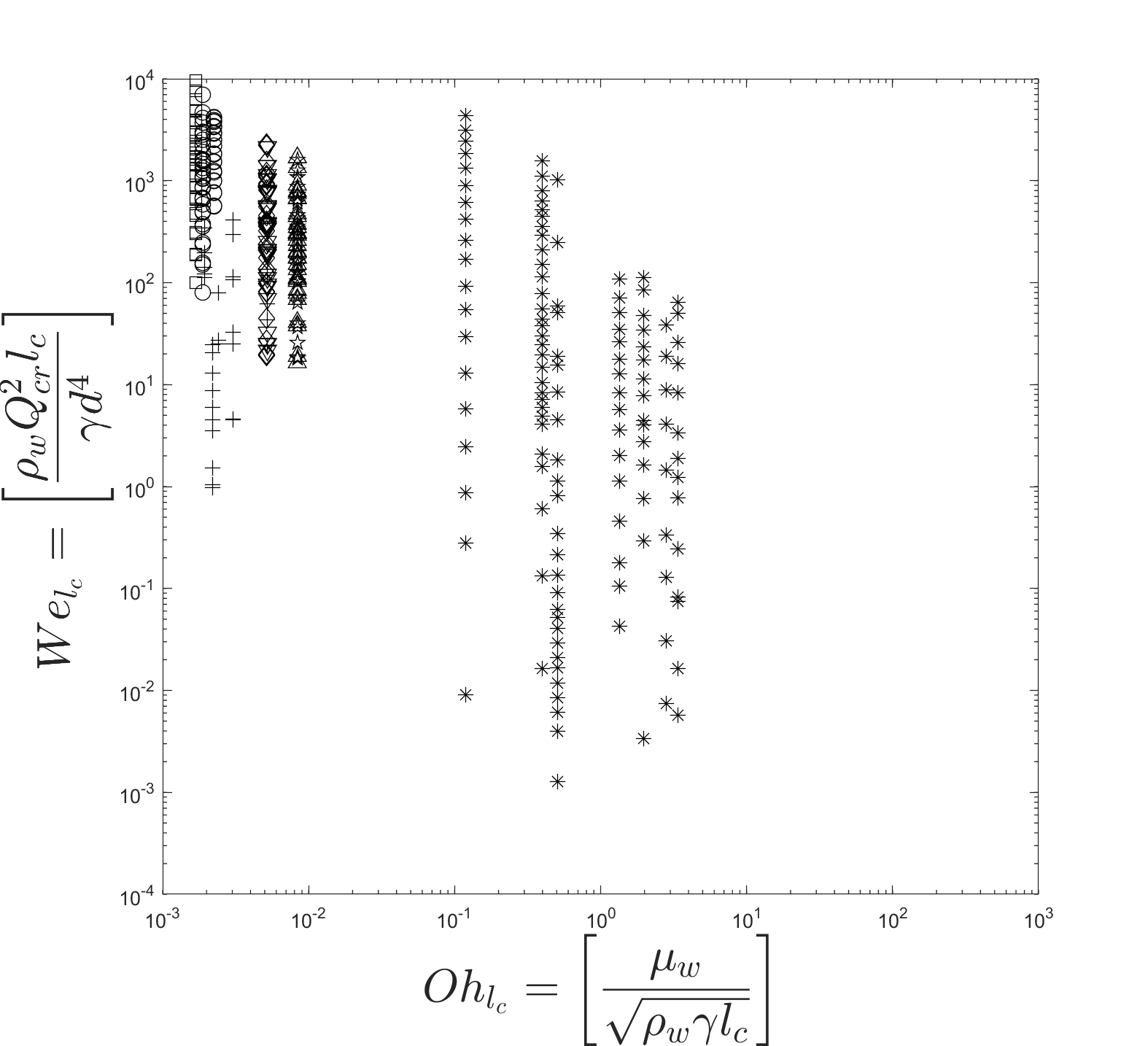}
\caption{\label{Ohn_regime}Weber number as a function of Ohnesorge number. The data from Cohen's experiment is shown using the symbol  \scrossvh. The data from Lubin and Springer's experiment is shown using the symbol  +. The data from this experiment had two orientations. The fluid combinations for the primary and the secondary orientation are expressed using the symbols mentioned in Table ~\ref{tab:table1}.}
\end{figure}

Initially, the relationships developed in the aforementioned works were applied to our data, but the results were not able to produce a consistent collapse for a predictive relationship between flow rate and submerged depth as shown in Fig.~\ref{appimage}.  In an attempt to understand the reason for the disparities, we consider the flow regime at the onset of entrainment using the Ohnesorge number ($Oh_{l_c}$), which is defined as the ratio of viscous effect to the combined effect of surface tension and inertia. As fluid is removed at a rate below critical flow rate, the reduction in the pressure due to the withdrawal flow is responsible for the deformation of the interface. In order to maintain the steady deformed interface the force balance has to be at an equilibrium. For a given sub-critical flow rate, at equilibrium, the flow regime effect is not explicitly observable other than causing a change in interface location or shape. But at the onset of entrainment, the interface is swept into the flow and the Ohnesorge number and Weber number help define the flow regime. Fig.~\ref{Ohn_regime} shows where the experimental studies fall in comparison to each other in terms of the Weber number and Ohnesorge number. There are three major regimes in terms of the Ohnesorge number. For $Oh_{l_c}<0.1$, the viscous effect is negligible compared to the surface tension and inertia effect. For $Oh_{l_c}>10$, the viscous effect dominates the surface tension and inertia effect. For $0.1<Oh_{l_c}<10$, the viscous effect is comparable relative to the surface tension and inertia effect. To understand the comparative effect between the surface tension and inertia Weber number was also plotted in  Fig.~\ref{Ohn_regime}. For $We_{l_c}>1$, the inertia effect dominates over surface tension and for $We_{l_c}<1$ the surface tension dominates inertia.

It was found that the Ohnesorge number for our experimental data was very low, meaning during entrainment the viscous effect is negligible compared to the surface tension effect and inertial effect, as shown in Fig.~\ref{Ohn_regime}. Similarly, Lubin's experimental data lies in the regime where viscous effects are also negligible. For both these studies the Weber numbers are high, meaning surface tension effects were not dominant. Reviewing Cohen's data in Fig.~\ref{Ohn_regime}, it can be seen that their experimental flow regime overlaps high and low values of $We_{l_c}$ with $Oh_{l_c}$ close to order one, that is some data are in the region where viscous effect was in a similar order in comparison to the surface tension and inertial effect and also in a region of surface tension dominance. This also explains the possibility of having a hysteresis effect at very low flow rates, as described by Cohen (\cite{cohen2004scaling}, \cite{cohen2002}).

In this paper, we present two approaches that were used to predict transition from selective withdrawal to entrainment for two withdrawal orientations -  a dimensional analysis based on the normal stress balance and an extension to a simplified inviscid model. No paper in the literature was found that studied withdrawal of bottom layer fluid from a tube penetrating the interface from above. Moreover, this analysis shows the comparison between withdrawal from above and withdrawal from below, especially due to the tube wall effect on the interface. Also it attempts to explain how having large diameter compared to the capillary length would affect the analysis, which would be more representative for real SPR-like conditions.

\begin{table*}
\caption{\label{tab:table1} Table of the properties for each fluid combination studied. $\rho_{u}$ and $\nu_{u}$ corresponds to upper fluid density and upper fluid kinematic viscosity respectively. $\rho_{w}$ and $\nu_{w}$ corresponds to lower fluid density and lower fluid kinematic viscosity respectively. $\gamma$ corresponds to surface tension coefficient. The upper layer fluid consists of two variations of polydimethyl siloxane (PDMS) and for lower fluid deionized water (DI $H_{2}O$) and two variations of Calcium Chloride Brine ($CaCl_{2}$ Brine) were used.}

\begin{ruledtabular}
\begin{tabular}{ccccccc}
         &System 1&System 2&System 3&System 4&System 5&System 6
 \\ \hline
 Symbols&\circlet&\trianglepb&\starlet&\squad&\rhombus&\trianglepa
 \\
 Upper Fluid&5 cSt PDMS&5 cSt PDMS&5 cSt PDMS&20 cSt PDMS&20 cSt PDMS&20 cSt PDMS
 \\
 Lower Fluid&DI $H_{2}O$&1.97 cSt $CaCl_{2}$ Brine&3 cSt $CaCl_{2}$ Brine&DI $H_{2}O$&1.97 cSt $CaCl_{2}$ Brine&3 cSt $CaCl_{2}$ Brine
 \\
 $\rho_{u} \; (g/cc) $&0.918&0.918&0.918&0.950&0.950&0.950
 \\
 $\rho_{w} \; (g/cc) $&0.998&1.245&1.325&0.998&1.245&1.325
 \\
 $\rho_{u}/\rho_{w}$&0.920&0.737&0.693&0.952&0.763&0.717
 \\
 $\nu_{u} \; (cSt) $&5&5&5&20&20&20
 \\
 $\nu_{w} \; (cSt) $&1.01&1.97&3&1.01&1.97&3
 \\
 $\nu_{u}/\nu_{w}$&4.95&2.54&1.67&19.80&10.15&6.67
 \\
 $\gamma \; (N/m)$&0.031&0.038&0.038&0.031&0.038&0.038
 \\
 $l_c/d$&0.44 - 0.94&0.22 - 0.46&0.20 - 0.41&0.57 - 1.21&0.23 - 0.49&0.20 - 0.43
\end{tabular}
\end{ruledtabular}
\end{table*}

\section{\label{sec:expdetail}Experimental Details\protect\\ }

A laboratory investigation was performed in order to gain insight into the conditions for which entrainment of the lighter fluid occurred for the orientations shown in Fig \ref{fig:1}. To record the transition from selective withdrawal to entrainment in an immiscible, two-liquid system, liquid pairs (silicon oils, brine, and water) were selected with density ratios similar to what was expected for SPR-like conditions. The liquid pairs were contained in a glass hexagonal tank approximately 25.4 cm wide and 61 cm tall. The withdrawal location was kept near the center of the container to minimize wall effects. Two withdrawal tube orientations were utilized. In the primary (SPR-like) orientation, the withdrawal tube is lowered through the interface and the lower fluid is drawn upwards (Fig.~\ref{fig:1}). In the secondary orientation, the withdrawal tube opening is placed below the interface and the lower fluid is drawn downwards. 

\begin{figure}
\includegraphics[width=86mm]{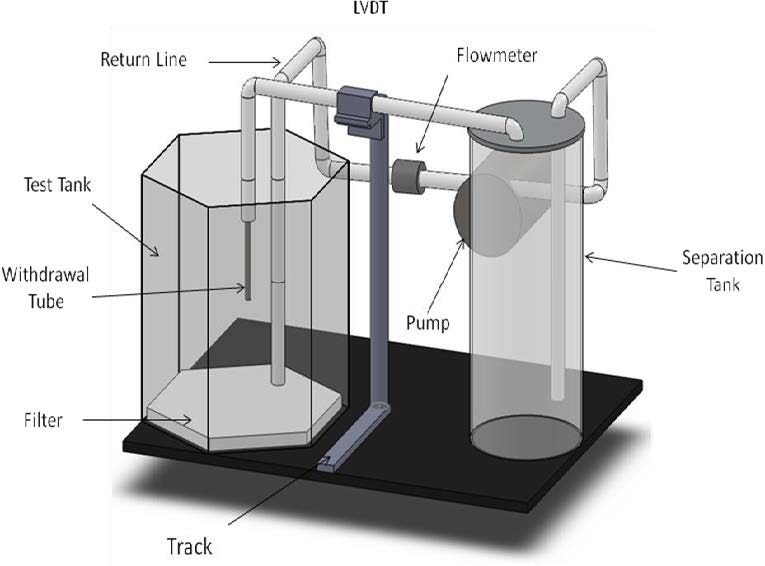}
\caption{\label{fig:2} Schematic diagram of the experimental setup. Cameras (not shown) were located perpendicular to the two faces of the hexagonal test tank.\cite{hartenberger2011}}
\end{figure}

All experiments began by adjusting the withdrawal tube centered in the tank such that the inlet was perpendicular to the undisturbed interface; two Redlake MotionPro cameras positioned 120$\degree$ apart were used to adjust alignment. The withdrawal tube was attached to a linear variable displacement transducer (lvdt) that was zeroed at the undisturbed interface. Prior to each experiment, the tube was lowered at least 2.54 cm below the interface of the upper fluid. The position of the tube was monitored using an lvdt (Fig.~\ref{fig:2}). The upper fluid layer was 2.54 cm thick. 

In order to ensure that the return flow from the filter at the bottom of the test tank did not significantly influence the flow at the liquid-liquid interface or near the withdrawal tube, dye injection tests were conducted. The results showed that the filter distributor created a flow uniform enough to maintain the flow quality. Furthermore, tests were conducted to show that there was no change on the critical submergence depth when the location of the withdrawal tube was varied by two diameters off-center or when the thickness of the oil layer was varied between 1.2 cm and 5.0 cm. This result is in contrast with the findings of Cohen \cite{cohen2004scaling} for selective entrainment above the interface, who reported that the entrainment process was affected by upper layer thickness less than 2.54 cm. 

After setting the initial position of the withdrawal tube, the withdrawal rate was then slowly increased to the desired level using an impeller pump. The lower fluid was pumped into a settling tank and then back into the bottom of the tank at the withdrawal rate, maintaining a constant interface depth. Keeping the flow rate constant, the tube was raised in increments of 0.25 mm until the interface neared transition at which time the tube was raised in increments of 0.03 mm until entrainment occurred. Upon entrainment the pump was stopped and the flow rate and tube depth were recorded. Six fluid pairs were tested in the experiments with two PDMS oils used as upper fluids and three brine solutions with varying concentrations of CaCl$_2$ as lower fluids. These combinations resulted in kinematic viscosity ratios from 1.67 to 19.80 and density ratios from 0.69 to 0.95. The surface tension for each pair was measured using a ring tensiometer and was between 0.03 N/m and 0.038 N/m for all pairs (Table~\ref{tab:table1}).

\begin{table} 
\caption{\label{tab:table2} Withdrawal tube dimensions. All the tubes used in the experiment were stainless-steel. }
\begin{ruledtabular}
\begin{tabular}{cccc}
        &Tube 1&Tube 2&Tube 3
 \\ \hline
 Nominal Dia. (cm)&0.953&1.27&1.905
 \\
 Inner Dia. (cm)&0.744&1.08&1.57
 \\
\end{tabular}
\end{ruledtabular}
\end{table}

For the primary orientation, three stainless steel tubes were used, with inner diameters of 0.74 cm, 1.08 cm, and 1.57 cm; the respective outer diameters were 0.95 cm, 1.27 cm, and 1.91 cm (Table~\ref{tab:table2}). The secondary orientation only utilized a stainless steel tube with an inner diameter of 1.08 cm. Flow rates were varied from $4.7 \times 10^{-5}$ m$^{3}/s$ to $6.5 \times 10^{-4}$ m$^{3}/s$ resulting in Reynolds number ($Re_{d}$) based on the inner tube diameter ranging from 2000 to 60,000, with most cases above $10^4$. Typical SPR-like situation has nominally 16,000 m$^{3}$/day of brine flowing  through  a  9.85 in. inner diameter tube. These flow conditions result in $Re_{d}$ values on the order of 390,000 to 900,000 \cite{hartenberger2011}. Assuming turbulent flow begins at $Re_{d}$ values on the order of 2000 and fully turbulent at $10^4$, the tube flow in the SPR caverns can be considered fully turbulent, similar to most of the experimental conditions presented.

\section{\label{sec:anal}Analysis \protect\\ }
 The first approach applies dimensional analysis on the normal stress balance equation. This analysis presents an idea about the forces which are important in the selective withdrawal phenomenon. The second approach uses Bernoulli's principle to explain the intuition achieved from the dimensional analysis pertaining to the experiment.

\subsection{\label{sec:dimanal}Interfacial Stress Balance}

The objective of this study was to determine a predictive relationship between the submergence depth and critical flowrate at the moment of entrainment. The dimensional analysis approach on the interfacial stress balance equation allows us to scrutinize the forces acting on the interface at selective withdrawal.
Surface tension manifests itself in the normal stress balance in the boundary condition for a Newtonian incompressible fluid as given below from \cite{deen2011}, where $\Delta P$ indicates the pressure change across the interface, followed by the viscous normal stress across the interface and the surface tension stress, where $\gamma$ is the surface tension and $\bf \nabla$ is the gradient operator.

\begin{eqnarray} \label{normalstress}
    \Delta P - \Delta \left(2\mu \frac{\partial  u_n}{\partial \bf n} \right)-\gamma \bf\nabla \cdot \bf n = 0
\label{normal}
\end{eqnarray}

Next, we nondimensionalize Eq.~(\ref{normalstress}) with a choice to make regarding the pressure scaling. In previous works, the flow was often considered to be creeping flow \cite{cohen2004scaling, lister1989, berkenbusch2008}, which resulted in a capillary number based on the capillary length as the representative nondimensional flow parameter that did not collapse our data. We therefore investigated dynamic pressure scaling and chose to bring gravity in the normal stress balance, by decomposing the pressure term into its dynamic and static component such that $P = P_d - \rho g z$, where, $P_d$ is the dynamic pressure, $g$ is gravitational acceleration and $z$ is the vertical height, defined positive opposing the gravity vector, thus $\vec{g} \cdot z$ is negative. Using $\rho_w U_0^2$ as the dynamic pressure scale, capillary length ($l_c = \sqrt{\gamma / \Delta \rho g}$) as length scale for the normal and tangential direction on the interface, tube diameter ($d$) as the axial or vertical length scale and $U_0 = 4Q_{cr}/\pi d^2 $ as the velocity scale respectively we achieve the following equation. The dimensionless variables are indicated by over-bars.

\begin{eqnarray}
    \overline{\Delta P_d} 
    - \frac{1}{Fr_{d}^2} \overline{\Delta \rho g z} -  \frac{1}{Re_{l_c}} \overline{\Delta \left(2\mu \frac{\partial  u_n}{\partial \bf n} \right)} \nonumber
    \\ 
    - \frac{1}{We_{l_c}} \overline{\gamma \bf \nabla \cdot \bf n} &=& 0
\label{normal_inviscid}
\end{eqnarray}

\noindent where, $Fr_{d} = U_0 / \sqrt{g' d}$, is the Froude number based on the tube diameter and $g'$ is the reduced gravity such that $g' = (1 - \rho_u / \rho_w)$, $\rho_u$ and $\rho_w$ are the upper fluid density and lower fluid density respectively. $Re_{l_c} = \rho_w U_0 l_c / \mu_w$ is the Reynolds number based on the capillary length and $We_{l_c} = \rho_w U_0^2 l_c/\gamma_s$ is the Weber number based on the capillary length.

\textit{Prior to entrainment, the location and shape of the interface is stable for a fixed flow rate, as observed from the experiment.} 
\textit{Using order of magnitude analysis, just at the onset of entrainment, it can be shown that shear term is couple of orders magnitude smaller.(See details in Appendix A)}

\begin{eqnarray}
    \underbrace{\overline{\Delta P_d}}_{O(1)}
    - \underbrace{\frac{1}{Fr_{d}^2} \overline{\Delta \rho g z}}_{O(10^{-1})} -  \underbrace{\frac{1}{Re_{l_c}} \overline{\Delta \left(2\mu \frac{\partial  u_n}{\partial \bf n} \right)}}_{O(10^{-2})} \nonumber
    \\ 
    - \underbrace{\frac{1}{We_{l_c}} \overline{\gamma \bf \nabla \cdot \bf n}}_{O(1)} &=& 0
\label{normal_inviscid_order}
\end{eqnarray}

\noindent\textit{The scaling analysis allowed the viscous shear component to be ignored and the dynamic pressure was found to be mostly balancing surface tension with a secondary effect of buoyancy.}

\textit{We defined the flow regime using the Ohnesorge number, $Oh_{l_c} = ( We_{l_c}^{\frac{1}{2}} / Re_{l_c} )$, as illustrated in Fig.~\ref{Ohn_regime}. This allows to Rearranging the Eq.~(\ref{normal_inviscid_order}) we can write the following,}

\begin{eqnarray}
    We_{l_c}^{\frac{1}{2}} \overline{\Delta P_d} 
    - \frac{We_{l_c}^{\frac{1}{2}}}{Fr_{d}^2} \overline{\Delta \rho g z} -  Oh_{l_c} \overline{\Delta \left(2\mu \frac{\partial  u_n}{\partial \bf n} \right)} \nonumber
    \\ 
    - \frac{1}{We_{l_c}^{\frac{1}{2}}} \overline{\gamma \bf \nabla \cdot \bf n} &=& 0
\label{normal_Ohn}
\end{eqnarray}

\textit{\noindent The Ohnesorge number has previously been used in the literature to define the droplet breakout regimes of liquid jets \cite{Eggers_2008}.} \textit{Recently, it has also been used to analyze the regimes of selective withdrawal phenomena \cite{panregime2020}. It enables the comparison of viscous effect to the surface tension and inertia effect in one nondimensional parameter. }
\textit{In our experiment the flow regime $Oh_{l_c}<0.0085$ and also from Eq.~(\ref{normal_inviscid_order}), it can be inferred that the viscous effect at the onset of entertainment is negligible for our experiments. This reduces the Eq.~(\ref{normal_Ohn}) into the following Eq.~(\ref{normal_final03}) as the representative stress balance equation.}

\begin{eqnarray}
    \overline{\Delta P_d} 
    - \frac{1}{We_{l_c}} \bigg[ \frac{1}{(l_c/d)} \times \overline{ \Delta \rho g z}
    + \overline{\gamma \bf \nabla \cdot \bf n}\bigg] &=& 0
\label{normal_final03}
\end{eqnarray}

There are three potential cases that can be considered in Eq.~(\ref{normal_final03}). For case 1, as $(l_c / d) \rightarrow \infty$, the surface tension balances the pressure and Weber number becomes dominant. For case 2, as $(l_c / d) \approx 1$, both buoyancy and the surface tension act as the resistive force to balance the pressure. But as the capillary length and the tube diameter are of similar order, the Weber number once again becomes the representative nondimensional parameter. For case 3, as the ratio $(l_c / d) \rightarrow 0$, the buoyancy effect balances the pressure term and Froude number becomes the dominant nondimensional parameter, which is the case for actual SPR.

Thus, dimensional analysis for our case indicates that the Weber number, based on capillary length,  is the representative flow parameter for the experimental data presented in this paper and is found to collapse the data well as shown in Fig.~\ref{weber02}.


\subsection{\label{sec:BernApprch}Inviscid Flow Model using Bernoulli's Principle}

This approach was introduced by Lubin and Springer \cite{lubin} and offered a more detailed insight on the selective withdrawal phenomena. The analysis was based on the assumptions that the viscous effects are negligible and the flow is incompressible. Bernoulli's principle was applied just below the interface streamline. Lubin and Springer, in their paper, assumed a hemispherical control volume surface area, with the plug hole being at the center of the hemisphere.

\begin{figure*}
\includegraphics[width=160 mm]{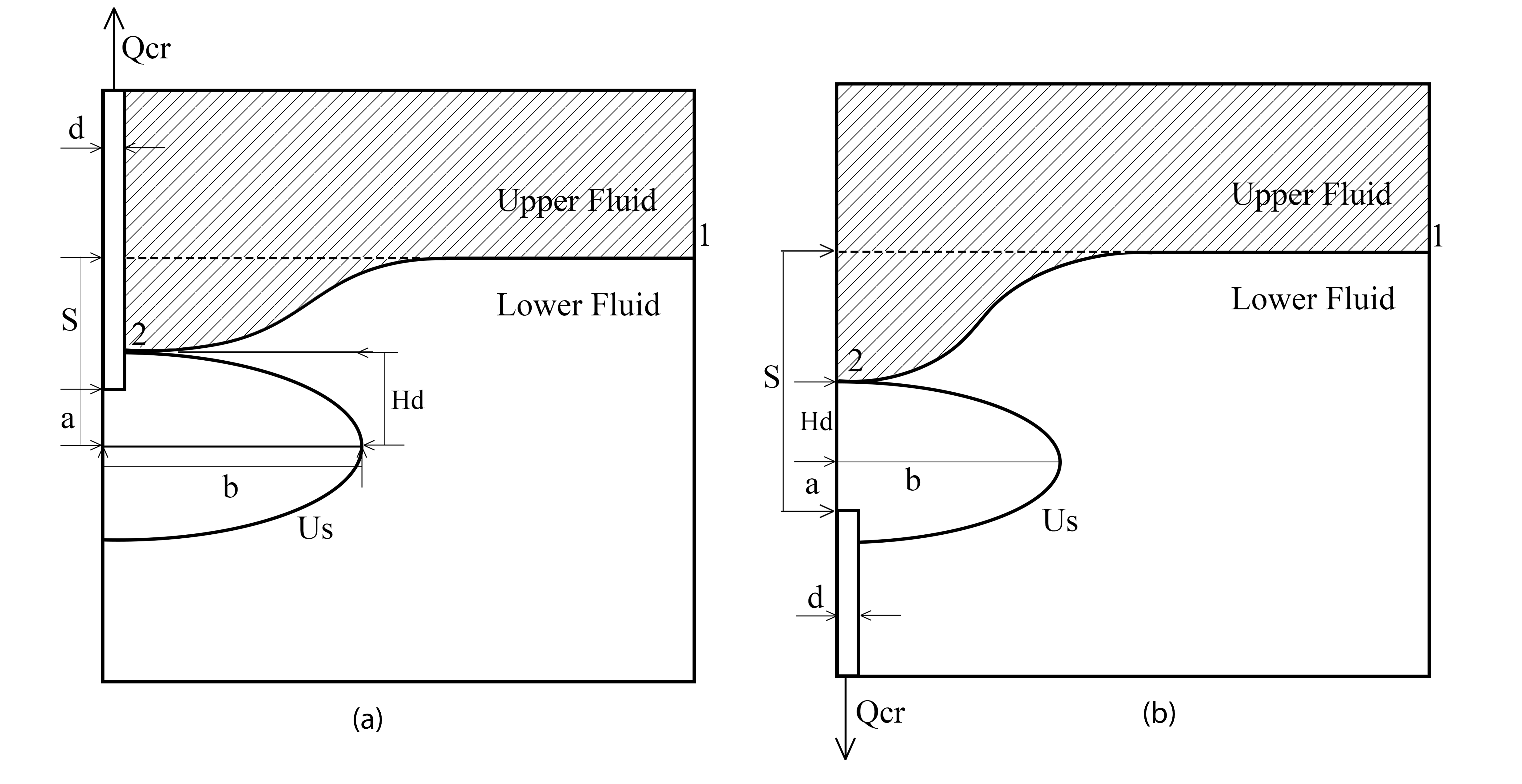}
 \caption{\label{schem_lubin01}Schematic of symbols and control volume used in the analysis of inviscid model. (a) Primary orientation (b) Secondary orientation. The critical flow rate is denoted by $Q_{cr}$, the tube diameter by $d$ and the submergence depth by $S$. $H_d$ and $b$ are the axes of the 2D ellipsoid. $U_s$ is the velocity normal to the control volume surface and $a$ is the offset of the ellipsoidal control volume from the tube opening.}
\end{figure*}




In this analysis, it was already confirmed, based on the Ohnesorge number, that the viscous effect was negligible for our experimental flow field, at the onset of entrainment. The assumption made by Lubin and Springer were slightly modified such that the model accounts for the suction tube instead of a drain-hole on a flat tank surface. An ellipsoidal control volume surface was assumed, with the center being at an offset ($a$) from the tube opening, as in Fig.~\ref{schem_lubin01}. Our supposition was that the interface acts similar to a static wall and influences the iso-velocity profiles away from the tube exit. An ellipsoidal control volume surface is more generalized as compared to the previously assumed spherical or hemispherical control volumes. This assumption is also supported by True and Crimaldi \cite{true2017} for inhalant flows for which velocity magnitude contours near the tube opening become ellipsoidal in shape as Re increased. The ellipsoidal shape is also see in mulitiphase Eulerian-Eulerian CFD simulations that we have performed for a few of our experimental cases. 

The flow phenomena observed in this experiment is similar to the one observed by Lubin and Springer \cite{lubin} and Hocking et al \cite{hocking_surfacetension_2016,farrow_hocking_2006}. At a certain entrainment depth and at a certain flow rate a dip forms above the point sink on the surface of the lower fluid. The flow is steady unless it reaches a critical entrainment depth ($S$) and a critical flow rate ($Q_{cr}$), at which point, the dip grows rapidly and extends towards the point sink almost instantaneously. Mass conservation at the ellipsoidal control volume, using Knudsen Thomsen approximation \cite{ellipse02}, can be expressed as,

\begin{eqnarray}
     Q_{cr} &=& 4 \pi \alpha_0 H_d^2 U_s 
\label{mass_flw_rte}
    \\
    \alpha_0 &\approx&  \left(\frac{b}{H_{d}}\right) \left[\frac{2}{3} \bigg\{1 + \frac{1}{2} \left(\frac{b}{H_d} \right)^{(\frac{8}{5})}\bigg\} \right]^{(\frac{5}{8})}
\label{defrmtn_coeff}
\end{eqnarray}

\noindent where, $Q_{cr}$ is the critical flow rate, $H_d$ and $b$ are the axes of the 2D ellipsoidal control surface. $U_s$ is average velocity normal to the ellipsoidal surface. The deformation coefficient, $\alpha_0$ expresses the deviation from the spherical control surface assumption. When $\alpha_0 \rightarrow 1$, the control surface is a sphere, when $\alpha_0 \rightarrow 0.5$, the control surface is a hemisphere, otherwise it is an ellipsoid. The $\alpha_0$ can be estimated using Eq.~(\ref{defrmtn_coeff}). 

Bernoulli's principle along a streamline from point 1 to point 2 in Fig.~\ref{schem_lubin01} generates the following equation,

\begin{eqnarray}
      S &=& H_d + \frac{U_s^2 }{2g'} \mp a
\label{pred_basic}
\end{eqnarray}

\noindent where, $(-)$ is for primary orientation, $(+)$ is for secondary orientation and $a$ is the offset of the ellipsoidal centroid from the tube opening. Eq.~(\ref{mass_flw_rte}) - Eq.~(\ref{pred_basic}) can be rearranged to yield,

\begin{eqnarray}
      S &=& H_d + \frac{Q_{cr}^2 }{32 \pi^2 g'\alpha_0^2 H_d^4} \mp a
\label{pred_med}
\end{eqnarray}

The assumption of instantaneous rupture of the interface at the critical condition is used to eliminate $H_d$.

\begin{eqnarray}
     \frac{dS}{dt} \big/ \frac{dH_d}{dt} \approx 0
\label{rupture}
\end{eqnarray}

\noindent Differentiating Eq.~(\ref{pred_med}) and applying Eq.~(\ref{rupture}) gives 

\begin{eqnarray}
    H_{d} &=& \frac{0.4174}{\alpha_0^{2/5}} \bigg[ \frac{Q_{cr}^2 }{g'}\bigg]^{\frac{1}{5}}
\label{curve_tip_dist}  
\end{eqnarray}

\noindent Substituting $H_d$ from Eq.~(\ref{curve_tip_dist}) to Eq.~(\ref{pred_med}) yields,

\begin{eqnarray}
     S &=& \frac{0.5227}{\alpha_0^{2/5}} \bigg[ \frac{Q_{cr}^2 }{g'}\bigg]^{\frac{1}{5}} \mp a
\label{pred_high}
\end{eqnarray}

Normalizing Eq.~(\ref{curve_tip_dist}) and Eq.~(\ref{pred_high}) with the diameter, allows us to re-write the equations in terms of Weber number, $We_{l_c} = \rho_w Q_{cr}^2 l_c/d^4\gamma$ and capillary length, $l_c = \sqrt{\gamma / \Delta \rho g}$, such that it provides a linear relationship between the nondimensionalized critical entrainment depth and nondimensionalized critical flow rate raised to the power one-fifth.

\begin{eqnarray}
     \frac{S}{d} &=& \frac{0.5227}{\alpha_0^{2/5}} \bigg( \frac{l_c}{d}\bigg)^{\frac{1}{5}} We_{l_c}^{\frac{1}{5}} \mp \frac{a}{d}
\label{pred_final}
\\
    \frac{H_{d}}{d} &=& \frac{0.4174}{\alpha_0^{2/5}}\bigg( \frac{l_c}{d}\bigg)^{\frac{1}{5}} We_{l_c}^{\frac{1}{5}}
\label{curve_tip_final}    
\end{eqnarray}

\begin{table}
\caption{\label{tab:table4} Fluid properties for calculating ratio of capillary length to tube diameter for Lubin's Paper \cite{lubin}}
\begin{ruledtabular}
\begin{tabular}{cccccc}
 $\rho_1(kg/m^3)$ &$\rho_2(kg/m^3)$&$\gamma(N/m)$&$d(m)$&$l_c(m)$&$(l_c/d)^{(1/5)}$
 \\ \hline
 998&1.2226&0.075&0.0032&0.003&0.99
 \\
 998&783.43&0.04&0.0032&0.004&1.05   
 \\
 998&918.16&0.021&0.0064&0.005&0.95
 \\
 998&918.16&0.038&0.0064&0.007&1.02
 \\
 998&868.26&0.012&0.0032&0.003&0.99
 \\
\end{tabular}
\end{ruledtabular}
\end{table}

\section{\label{sec:disc}Results and Discussion \protect\\ }

\begin{figure}
\includegraphics[width=86mm]{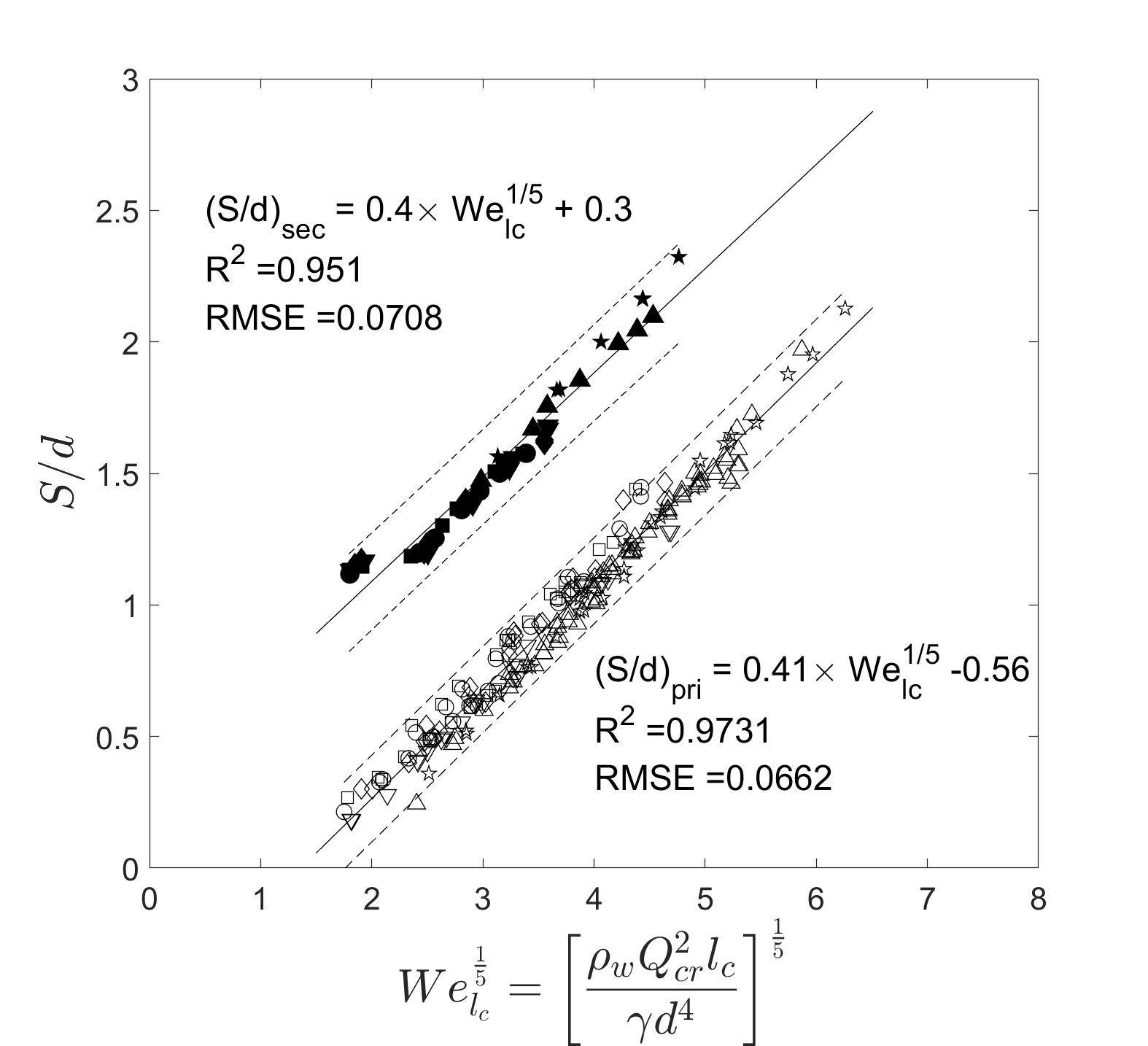}
\caption{\label{weber02}Critical submergence depth as a function of critical Weber number. The "hollow" symbols represent the primary orientation data, whereas the "filled" ones represent the secondary orientation data. The solid line through the "hollow" symbols is the best linear fit for the primary orientation and the solid line through the "filled" symbols is the best linear fit for the secondary orientation. The dotted lines are 95\% prediction intervals. The region above the 95\% upper bound of entrainment depth is selective withdrawal zone, whereas the area below is entrainment zone. The symbols are defined in Table ~\ref{tab:table1}.}
\end{figure}

We now apply Eq.~(\ref{pred_final}) to our experimental data for both orientations and present the results in Fig.~\ref{weber02}. The relation for entrainment depth as a function of Weber number to the one-fifth power fits the data well. Additionally, expressions for $\alpha_0$, $a$ and $b$ can easily be estimated from the slope and intercept values of the fitted equation as shown in Table \ref{tab:table3}.

\begin{table*}
\caption{\label{tab:table3} Table of estimations for the parameters  $\alpha_0$, $a$ and $b$}
\begin{ruledtabular}
\begin{tabular}{cccccc}
 Orientation&slope&intercept&$\alpha_0$&$a(m)$&$b(m)$
 \\ \hline
 Primary&0.41($\pm$ 0.01)&-0.56($\pm$ 0.04)&$1.84\big(l_c/d\big)^{\frac{1}{2}}$&$0.56d$&$H_{d}\big[ -1+ \big\{ 1 + 7.92  \big(l_c/d\big)^{\frac{4}{5}}\big\}^{\frac{1}{2}} \big]^{\frac{5}{8}}$
 \\
 Secondary&0.40($\pm$ 0.02)&+0.30($\pm$ 0.08)&$1.95\big(l_c/d\big)^{\frac{1}{2}}$&$0.30d$&$H_{d}\big[ -1+ \big\{ 1 + 8.75  \big(l_c/d\big)^{\frac{4}{5}}\big\}^{\frac{1}{2}} \big]^{\frac{5}{8}}$
 \\
\end{tabular}
\end{ruledtabular}
\end{table*}

From this analysis, it can also be concluded that the experiment by Lubin and Springer is a special case of the proposed model, where $\alpha_0 = 0.5$ (for hemispherical control volume surface) and the fluid layers were chosen such that $(l_c/d)^{1/5} \approx 1$, see Table \ref{tab:table4}. The point sink was assumed to be exactly on the plug-hole opening, rendering $a = 0$. Using these assumptions in Eq.~(\ref{mass_flw_rte}), Eq.~(\ref{curve_tip_dist}) and Eq.~(\ref{pred_high}), as well as taking help from Eq.~(\ref{eqn:Fr_We_final}) the expression derived by Lubin in Eq.~(\ref{lubin_corr}) is realized

\begin{eqnarray}
 \frac{S}{d} = 0.69 \left[\frac{Q_{cr}^2}{g'd^5} \right]^{\frac{1}{5}}
\label{lubin_corr}
\end{eqnarray}

For an individual critical Weber number, the fit equations, lead to a critical nondimensional submergence depth for the corresponding tube orientations, Fig.~\ref{weber02}. For cases when the viscous effect is negligible compared to the surface tension and inertial effect (low $Oh_{l_c}$ flow), and the capillary length is in the comparable order of scale to the tube diameter, a  predictive relationship is generated for the desired selective withdrawal, above the curve, and entrainment, below the curve.

The offset between the primary and the secondary orientation can be explained as the effect of the tube wall on the interface. In the primary orientation, the tube wall is in contact with the interface and impedes the entrainment whereas in the secondary orientation the interface is completely free from any surface contact. As a result, for the same submergence depth, primary orientation requires a stronger flow rate compared to the secondary orientation, for entrainment to occur. A point to be noted here is that having a negative entrainment depth would require the lower fluid to ``stick'' to the withdrawal tube and block the upper fluid from being entrained. 

It is evident from the analysis, that the ratio $(l_c/d)$ plays a significant role in shaping the ellipsoid, see Table~\ref{tab:table3} for parameters. The inverse Bond number can be defined as $(l_c/d)$, details in Appendix B.  It can be interpreted that $Bo^{-1}$ is the scaled surface tension effect with respect to the buoyancy effect at the diameter scale. From the normal stress balance  equation, Eq.~(\ref{normal_final03}), buoyancy dominates surface tension for large diameters and the $Bo^{-1}$ reduces to a small number. In SPR-like cavern flows, the tube diameter is large compared to the capillary length, so it can be inferred that in SPR-like conditions the force balance is one of buoyancy balancing pressure and consequently the Froude number is expected to be the representative nondimensional number, and the correlation shown in Fig. \ref{weber02} is not claimed to be predictive. However, for the experiments presented herein, the inverse Bo numbers range from about 0.2 to 1.2, which is in the vicinity of case 2 and makes Weber number the representative nondimensional number at entrainment.

\begin{figure}
\includegraphics[width=86mm]{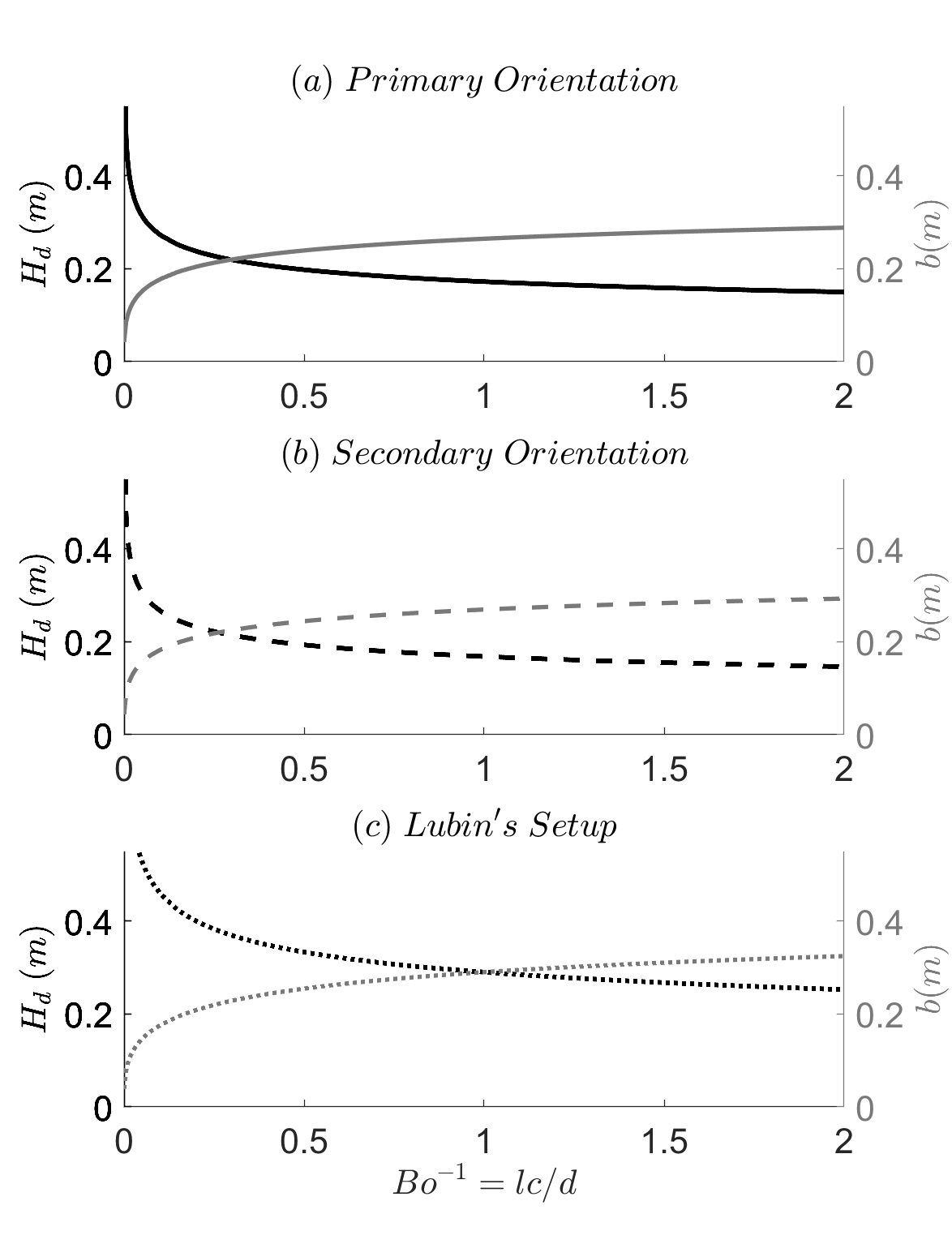}
\caption{\label{Hd_bvsBo_BnW}Ellipsoidal control volume parameters $(H_d,b)$ as a function of the inverse Bond number for (a) primary orientation (solid line), (b) secondary orientation (dashed line) and (c) Lubin and Springer model (dotted line). The fluid combination of system 3 (Table \ref{tab:table1}) was chosen. The flow rate was set to be 0.35 $m^3/s$. The surface tension coefficient was varied from 0.001 N/m to 0.1 N/m and eight selected diameters were chosen from a very wide range of 0.001 m to 0.30 m.}
\end{figure}

In this experiment, during selective withdrawal, the reduction of pressure due to the tube velocity creates a downward pull on the interface toward the tube inlet. The surface tension and the buoyancy create a resistance to this force and at equilibrium, a steady state, sub-critical balance is achieved such as shown in the left photo in Fig \ref{fig:1}. When $Bo^{-1} > 1$, surface tension dominates as the resistive force and when $Bo^{-1} < 1$, buoyancy dominates. Entrainment occurs when the resistive force is overcome by the force due to pressure reduction.

To understand the effect on inverse Bond number on the shape of the ellipsoidal control volume, a simulation was conducted keeping the flow rate fixed for a chosen fluid combination. A high flow rate of 0.35 $m^3/s$ was chosen so that the high Reynolds number flow condition was satisfied even for the largest diameter and viscous effect can be neglected ($Re_{l_c} \approx 10^3 \;to\; 10^8$). It corresponds to very low Ohnesorge number ($Oh_{l_c} \approx 10^{-4} \;to\; 10^{-2}$) for the simulation. The inverse Bond number, $Bo^{-1} = (l_c/d)$ was varied by changing the surface tension coefficient and the diameter. 
The range of the surface tension coefficient was from 0.001 N/m to 0.1 N/m. Eight selected diameters were chosen from a very wide range of 0.001 m to 0.30 m.

The inverse Bond number, according to the derived model, shapes the major and minor axes of the ellipsoidal control volume surface. Fig.~\ref{Hd_bvsBo_BnW}(a), shows the trend of $H_d,b$ for the primary orientation as a function of $Bo^{-1}$. The trend for the secondary orientation Fig.~\ref{Hd_bvsBo_BnW}(b) was found to be similar the primary. For both orientations, the ellipsoid becomes a sphere at a particular $Bo^{-1}$ value. As the $Bo^{-1}$ increases, $H_d$ tends to decrease, indicating the lowering of interface, Fig.~\ref{schem_lubin01}.  The interface acts like a static wall and consequently, as the interface drops, it also reduces the velocity in its vicinity. At some critical velocity $U_s$, the interface cannot resist the downward pull anymore and it collapses. To collapse at a reduced critical velocity the area of the iso-velocity surface has to increase, as the $Bo^{-1}$ increases. The only way the iso-velocity area can increase for a fixed flow rate is by an increase of the parameter $b$, which is evident in Fig.~\ref{Hd_bvsBo_BnW}. Figure~\ref{Hd_bvsBo_BnW}(c) shows the results of extending the analysis to fluids in Lubin's experimental setup. It is apparent that as $Bo^{-1}$ gets closer to 1, the major and minor axes becomes equal and the control volume shape obtains the form of an hemisphere, as Lubin and Spring assumed in his paper.

\section{\label{sec:conc}Conclusion \protect\\ }

This paper presented two approaches to predict transition from selective withdrawal to entrainment using the physics of fluid flow. The first used the dimensional analysis approach on the normal stress balance equation showed the Weber number as the relevant nondimensional parameter specific to selective withdrawal. Moreover, this method provided a representation for the force balance on the interface for the relative significance of buoyancy, surface tension, and pressure balance. The second used Bernoulli's principle for selective withdrawal with two withdrawal-tube orientations. The theoretical model was fitted using experimental data and the expressions for corresponding unknown parameters were derived. It is shown that the general control volume associated with surfaces of iso-velocity at the critical flow rate is ellipsoidal in shape. It was also shown that Lubin's correlation is a special case of this proposed model.  Both approaches led to predictive relations for the selective entrainment depth as a function of the critical Weber number raised to the one-fifth power.

\begin{acknowledgments}
The experimental work was performed at Sandia National Laboratories with Joel Hartenberger, Tim O'Hern, and Stephen Webb.
\end{acknowledgments}

\appendix

\section{}

Order of magnitude analysis for each term of Eq.~(\ref{normal_inviscid_order}) can be performed using the appropriate scales.

\noindent Analyzing $\overline{\Delta P_d}$ term:

\begin{eqnarray}
 \overline{\Delta P_d} = \frac{(\rho_w U_i^2 - \rho_u U_i^2)} {\rho_w U_0^2 } \approx O(\frac{U_i^2}{U_0^2})
\label{delPd_scale_01}
\end{eqnarray}

\noindent where $U_i (\approx Q_{cr}/4\pi l_c^2)$ is the velocity estimated close to the interface. The maximum order of the ratio $U_i/U_0$ becomes $10^{-1}$.

\begin{eqnarray}
 \overline{\Delta P_d} = \frac{(\rho_w U_i^2 - \rho_u U_i^2)} {\rho_w U_0^2 } \approx O(10^{-2})
\label{delPd_scale_02}
\end{eqnarray}

\noindent Analyzing $\frac{1}{Fr_d^2} \overline{\Delta \rho g z}$ term:

\begin{eqnarray}
 \overline{\Delta \rho g z} = \frac{(\rho_w g z_w - \rho_u g z_u)} {\rho_w g' d} \approx O(\frac{\rho_w g z_w}{\rho_w g' d})
\label{delrhogz_scale_01}
\end{eqnarray}

\noindent The maximum depth that z can have is in the order of the submergence depth S. So the maximum order becomes,

\begin{eqnarray}
 \overline{\Delta \rho g z} = \frac{(\rho_w g z_w - \rho_u g z_u)} {\rho_w g' d} \approx O(1)
\label{delrhogz_scale_02}
\end{eqnarray}

\noindent Similarly, for the Froude number we can find the maximum order to be,

\begin{eqnarray}
 \frac{1}{Fr_d^2} = \frac{g'd}{U_0^2} \approx \frac{10 \times 10^{-2}}{10^2} \approx O(10^{-3})
\label{delrhogz_scale_03}
\end{eqnarray}

\noindent The order of the total product becomes,

\begin{eqnarray}
 \frac{1}{Fr_d^2} \overline{\Delta \rho g z} \approx O(10^{-3})
\label{delrhogz_scale_04}
\end{eqnarray}

\noindent Analyzing $\frac{1}{Re_{l_c}} \overline{\Delta \left(2\mu \frac{\partial  u_n}{\partial n} \right)}$ term:

\begin{eqnarray}
 \overline{\Delta \left(\mu \frac{\partial  u_n}{\partial n} \right)} = \frac{\mu_w \frac{\Delta u_n^w}{\Delta n} - \mu_w \frac{\Delta u_n^u}{\Delta n}}{\mu_w \frac{U_0}{l_c}}
\label{delvisc_scale_01}
\end{eqnarray}

\noindent Below the interface, for the withdrawal fluid, it can be estimated that the velocity goes from almost zero (compared to tube velocity) near the interface to $U_0$ near the tube, over the length scale $l_c$. Above the interface, for the upper fluid, it can be estimated that the velocity goes from almost zero near the interface to some maximum normal velocity $U_n^m (\approx 4Q_{cr}/\pi l_c^2)$ over the length scale $l_c$. Using these estimated scales we can rewrite Eq.~(\ref{delvisc_scale_01}) as below,

\begin{eqnarray}
 \overline{\Delta \left(\mu \frac{\partial  u_n}{\partial n} \right)} \approx O(\frac{\mu_u U_n^m}{\mu_w U_0}) \approx O(10)
\label{delvisc_scale_02}
\end{eqnarray}

\noindent Similarly, for the Reynolds number we can find the order to be,

\begin{eqnarray}
 \frac{1}{Re_{l_c}} = \frac{\mu_w}{\rho_w U_0 l_c} \approx \frac{10^{-3}}{10^3 \times 10 \times 10^{-2}} \approx O(10^{-5})
\label{delvisc_scale_03}
\end{eqnarray}

\noindent The order of the product becomes,

\begin{eqnarray}
 \frac{1}{Re_{l_c}} \overline{\Delta \left(2\mu \frac{\partial  u_n}{\partial n} \right)} \approx O(10^{-4})
\label{delvisc_scale_04}
\end{eqnarray}

\noindent Analyzing $\frac{1}{We_{l_c}}\overline{\gamma \bf \nabla \cdot \bf n}$ term:

\begin{eqnarray}
 \overline{\gamma \bf \nabla \cdot \bf n} = \frac{\gamma \bf \nabla \cdot \bf n }{\gamma_s l_c} \approx O(\frac{\bf \nabla \cdot \bf n}{l_c} )
\label{delgamma_scale_01}
\end{eqnarray}

\noindent where $\bf \nabla \cdot \bf n$ is the curvature and from our experiment it is in the order of 10.  

\begin{eqnarray}
 \overline{\gamma \bf \nabla \cdot \bf n} \approx O(\frac{\bf \nabla \cdot \bf n}{l_c} ) \approx O(10^3)
\label{delgamma_scale_02}
\end{eqnarray}

\noindent The order of Weber number in our experiments are,

\begin{eqnarray}
 \frac{1}{We_{l_c}} = \frac{\gamma_s}{\rho_w U_0^2 l_c} \approx \frac{10^{-2}}{10^3 . 10^2 .10^{-2}} \approx O(10^{-5})
\label{delgamma_scale_03}
\end{eqnarray}

\noindent The order of the product is as following,

\begin{eqnarray}
 \frac{1}{We_{l_c}}\overline{\gamma \bf \nabla \cdot \bf n} = \approx O(10^{-2})
\label{delgamma_scale_04}
\end{eqnarray}

\section{}

The Froude number for two-fluid stratified flow is defined as the ratio of flow inertia to buoyancy effect \cite{white2016,smoke_TANG2017}, $Fr_{L_1} = U_0 / \sqrt{g' L_1}$, where $g'$ is the reduced gravity such that $g' = (1 - \rho_u / \rho_w)$, assuming $\rho_w > \rho_u$. Similarly, the ratio of inertia to surface tension effect is defined as Weber number \cite{white2016}, $We_{L_2} = (\rho_w U_0^2) / (\gamma_s/L_2)$. When comparing Froude number to the Weber number, another dimensionless number, known as the Bond number \cite{white2016}, comes into effect. The Bond number is defined as the ratio of buoyancy effect to surface tension effect, $Bo = (\rho_w g' L_1) / (\gamma_s/L_2)$. The squared Froude number can be re-written as,

\begin{eqnarray}
    Fr_{L_1}^2 &=& \frac{U_0^2}{g'L_1} = \frac{\rho_w U_0^2 L_2}{\gamma_s} \times \frac{\gamma_s}{\rho_w g'L_1L_2}
\label{eqn:Fr_We_raw}
\\
    Fr_{L_1}^2 &=& We_{L_2}/Bo
\label{eqn:Fr_We_semifinal}    
\end{eqnarray}

The dimensional analysis approach uses length scales $L_1 = L_2 = l_c$. Using the definition of capillary length, $l_c$, the Bond number reduces to $Bo = 1$. Applying in Eq.~(\ref{eqn:Fr_We_semifinal}) yields,

\begin{eqnarray}
   Fr_{l_c}^2 &=& We_{l_c} 
\label{eqn:Fr_We_lc}    
\end{eqnarray}

Similarly, if the length scales were such that $L_1 = d$ and $L_2 = l_c$, the Bond number reduces to $Bo = d / l_c$. Applying in Eq.~(\ref{eqn:Fr_We_semifinal}) yields,

\begin{eqnarray}
   Fr_{d}^2 &=& We_{l_c}/Bo = \bigg( \frac{l_c}{d}\bigg)We_{l_c}
\label{eqn:Fr_We_final}    
\end{eqnarray}

Eq.~(\ref{eqn:Fr_We_lc}) and Eq.~(\ref{eqn:Fr_We_final}) helps us express the Froude numbers in terms of Weber numbers.



\providecommand{\noopsort}[1]{}\providecommand{\singleletter}[1]{#1}%
\begin{thebibliography}{17}%
\makeatletter
\providecommand \@ifxundefined [1]{%
 \@ifx{#1\undefined}
}%
\providecommand \@ifnum [1]{%
 \ifnum #1\expandafter \@firstoftwo
 \else \expandafter \@secondoftwo
 \fi
}%
\providecommand \@ifx [1]{%
 \ifx #1\expandafter \@firstoftwo
 \else \expandafter \@secondoftwo
 \fi
}%
\providecommand \natexlab [1]{#1}%
\providecommand \enquote  [1]{``#1''}%
\providecommand \bibnamefont  [1]{#1}%
\providecommand \bibfnamefont [1]{#1}%
\providecommand \citenamefont [1]{#1}%
\providecommand \href@noop [0]{\@secondoftwo}%
\providecommand \href [0]{\begingroup \@sanitize@url \@href}%
\providecommand \@href[1]{\@@startlink{#1}\@@href}%
\providecommand \@@href[1]{\endgroup#1\@@endlink}%
\providecommand \@sanitize@url [0]{\catcode `\\12\catcode `\$12\catcode
  `\&12\catcode `\#12\catcode `\^12\catcode `\_12\catcode `\%12\relax}%
\providecommand \@@startlink[1]{}%
\providecommand \@@endlink[0]{}%
\providecommand \url  [0]{\begingroup\@sanitize@url \@url }%
\providecommand \@url [1]{\endgroup\@href {#1}{\urlprefix }}%
\providecommand \urlprefix  [0]{URL }%
\providecommand \Eprint [0]{\href }%
\providecommand \doibase [0]{https://doi.org/}%
\providecommand \selectlanguage [0]{\@gobble}%
\providecommand \bibinfo  [0]{\@secondoftwo}%
\providecommand \bibfield  [0]{\@secondoftwo}%
\providecommand \translation [1]{[#1]}%
\providecommand \BibitemOpen [0]{}%
\providecommand \bibitemStop [0]{}%
\providecommand \bibitemNoStop [0]{.\EOS\space}%
\providecommand \EOS [0]{\spacefactor3000\relax}%
\providecommand \BibitemShut  [1]{\csname bibitem#1\endcsname}%
\let\auto@bib@innerbib\@empty
\bibitem [{\citenamefont {Hartenberger}\ and\ \citenamefont
  {O’Hern}(2011)}]{hartenberger2011}%
  \BibitemOpen
  \bibfield  {author} {\bibinfo {author} {\bibfnamefont {J.~D.}\ \bibnamefont
  {Hartenberger}}\ and\ \bibinfo {author} {\bibfnamefont {T.~J.}\ \bibnamefont
  {O’Hern}}\ }(\bibinfo {year} {2011})\ pp.\ \bibinfo {pages}
  {1299--1307}\BibitemShut {NoStop}%
\bibitem [{\citenamefont {Lubin}\ and\ \citenamefont {Springer}(1967)}]{lubin}%
  \BibitemOpen
  \bibfield  {author} {\bibinfo {author} {\bibfnamefont {B.~T.}\ \bibnamefont
  {Lubin}}\ and\ \bibinfo {author} {\bibfnamefont {G.}~\bibnamefont
  {Springer}},\ }\href {https://doi.org/10.1017/S0022112067000898} {\bibfield
  {journal} {\bibinfo  {journal} {Journal of Fluid Mechanics}\ }\textbf
  {\bibinfo {volume} {29}},\ \bibinfo {pages} {385–390} (\bibinfo {year}
  {1967})}\BibitemShut {NoStop}%
\bibitem [{\citenamefont {Cohen}\ and\ \citenamefont
  {Nagel}(2002)}]{cohen2002}%
  \BibitemOpen
  \bibfield  {author} {\bibinfo {author} {\bibfnamefont {I.}~\bibnamefont
  {Cohen}}\ and\ \bibinfo {author} {\bibfnamefont {S.~R.}\ \bibnamefont
  {Nagel}},\ }\href {https://doi.org/10.1103/PhysRevLett.88.074501} {\bibfield
  {journal} {\bibinfo  {journal} {Phys. Rev. Lett.}\ }\textbf {\bibinfo
  {volume} {88}},\ \bibinfo {pages} {074501} (\bibinfo {year}
  {2002})}\BibitemShut {NoStop}%
\bibitem [{\citenamefont {Cohen}(2004)}]{cohen2004scaling}%
  \BibitemOpen
  \bibfield  {author} {\bibinfo {author} {\bibfnamefont {I.}~\bibnamefont
  {Cohen}},\ }\href {https://doi.org/10.1103/PhysRevE.70.026302} {\bibfield
  {journal} {\bibinfo  {journal} {Phys. Rev. E}\ }\textbf {\bibinfo {volume}
  {70}},\ \bibinfo {pages} {026302} (\bibinfo {year} {2004})}\BibitemShut
  {NoStop}%
\bibitem [{\citenamefont {Case}\ and\ \citenamefont {Nagel}(2007)}]{case2007}%
  \BibitemOpen
  \bibfield  {author} {\bibinfo {author} {\bibfnamefont {S.~C.}\ \bibnamefont
  {Case}}\ and\ \bibinfo {author} {\bibfnamefont {S.~R.}\ \bibnamefont
  {Nagel}},\ }\href {https://doi.org/10.1103/PhysRevLett.98.114501} {\bibfield
  {journal} {\bibinfo  {journal} {Phys. Rev. Lett.}\ }\textbf {\bibinfo
  {volume} {98}},\ \bibinfo {pages} {114501} (\bibinfo {year}
  {2007})}\BibitemShut {NoStop}%
\bibitem [{\citenamefont {Blanchette}\ and\ \citenamefont
  {Zhang}(2009)}]{blanchette2009}%
  \BibitemOpen
  \bibfield  {author} {\bibinfo {author} {\bibfnamefont {F.~m.~c.}\
  \bibnamefont {Blanchette}}\ and\ \bibinfo {author} {\bibfnamefont {W.~W.}\
  \bibnamefont {Zhang}},\ }\href
  {https://doi.org/10.1103/PhysRevLett.102.144501} {\bibfield  {journal}
  {\bibinfo  {journal} {Phys. Rev. Lett.}\ }\textbf {\bibinfo {volume} {102}},\
  \bibinfo {pages} {144501} (\bibinfo {year} {2009})}\BibitemShut {NoStop}%
\bibitem [{\citenamefont {Lister}(1989)}]{lister1989}%
  \BibitemOpen
  \bibfield  {author} {\bibinfo {author} {\bibfnamefont {J.~R.}\ \bibnamefont
  {Lister}},\ }\href {https://doi.org/10.1017/S002211208900011X} {\bibfield
  {journal} {\bibinfo  {journal} {Journal of Fluid Mechanics}\ }\textbf
  {\bibinfo {volume} {198}},\ \bibinfo {pages} {231–254} (\bibinfo {year}
  {1989})}\BibitemShut {NoStop}%
\bibitem [{\citenamefont {Farrow}\ and\ \citenamefont
  {Hocking}(2006)}]{farrow_hocking_2006}%
  \BibitemOpen
  \bibfield  {author} {\bibinfo {author} {\bibfnamefont {D.~E.}\ \bibnamefont
  {Farrow}}\ and\ \bibinfo {author} {\bibfnamefont {G.~C.}\ \bibnamefont
  {Hocking}},\ }\href {https://doi.org/10.1017/S0022112005007561} {\bibfield
  {journal} {\bibinfo  {journal} {Journal of Fluid Mechanics}\ }\textbf
  {\bibinfo {volume} {549}},\ \bibinfo {pages} {141–157} (\bibinfo {year}
  {2006})}\BibitemShut {NoStop}%
\bibitem [{\citenamefont {Hocking}\ \emph {et~al.}(2016)\citenamefont
  {Hocking}, \citenamefont {Nguyen}, \citenamefont {Forbes},\ and\
  \citenamefont {Stokes}}]{hocking_surfacetension_2016}%
  \BibitemOpen
  \bibfield  {author} {\bibinfo {author} {\bibfnamefont {G.~C.}\ \bibnamefont
  {Hocking}}, \bibinfo {author} {\bibfnamefont {H.~H.~N.}\ \bibnamefont
  {Nguyen}}, \bibinfo {author} {\bibfnamefont {L.~K.}\ \bibnamefont {Forbes}},\
  and\ \bibinfo {author} {\bibfnamefont {T.~E.}\ \bibnamefont {Stokes}},\
  }\href {https://doi.org/10.1017/S1446181116000018} {\bibfield  {journal}
  {\bibinfo  {journal} {The ANZIAM Journal}\ }\textbf {\bibinfo {volume}
  {57}},\ \bibinfo {pages} {417–428} (\bibinfo {year} {2016})}\BibitemShut
  {NoStop}%
\bibitem [{\citenamefont {Deen}(2011)}]{deen2011}%
  \BibitemOpen
  \bibfield  {author} {\bibinfo {author} {\bibfnamefont {W.~M.}\ \bibnamefont
  {Deen}},\ }\href@noop {} {\emph {\bibinfo {title} {Analysis of Transport
  Phenomena}}},\ Topics in Chemical Engineering\ (\bibinfo  {publisher} {Oxford
  University Press},\ \bibinfo {year} {2011})\BibitemShut {NoStop}%
\bibitem [{\citenamefont {Berkenbusch}\ \emph {et~al.}(2008)\citenamefont
  {Berkenbusch}, \citenamefont {Cohen},\ and\ \citenamefont
  {Zhang}}]{berkenbusch2008}%
  \BibitemOpen
  \bibfield  {author} {\bibinfo {author} {\bibfnamefont {M.~K.}\ \bibnamefont
  {Berkenbusch}}, \bibinfo {author} {\bibfnamefont {I.}~\bibnamefont {Cohen}},\
  and\ \bibinfo {author} {\bibfnamefont {W.~W.}\ \bibnamefont {Zhang}},\ }\href
  {https://doi.org/10.1017/S0022112008001900} {\bibfield  {journal} {\bibinfo
  {journal} {Journal of Fluid Mechanics}\ }\textbf {\bibinfo {volume} {613}},\
  \bibinfo {pages} {171–203} (\bibinfo {year} {2008})}\BibitemShut {NoStop}%
\bibitem [{\citenamefont {Eggers}\ and\ \citenamefont
  {Villermaux}(2008)}]{Eggers_2008}%
  \BibitemOpen
  \bibfield  {author} {\bibinfo {author} {\bibfnamefont {J.}~\bibnamefont
  {Eggers}}\ and\ \bibinfo {author} {\bibfnamefont {E.}~\bibnamefont
  {Villermaux}},\ }\href {https://doi.org/10.1088/0034-4885/71/3/036601}
  {\bibfield  {journal} {\bibinfo  {journal} {Reports on Progress in Physics}\
  }\textbf {\bibinfo {volume} {71}},\ \bibinfo {pages} {036601} (\bibinfo
  {year} {2008})}\BibitemShut {NoStop}%
\bibitem [{\citenamefont {Pan}\ \emph {et~al.}(2020)\citenamefont {Pan},
  \citenamefont {Nunes},\ and\ \citenamefont {Stone}}]{panregime2020}%
  \BibitemOpen
  \bibfield  {author} {\bibinfo {author} {\bibfnamefont {Z.}~\bibnamefont
  {Pan}}, \bibinfo {author} {\bibfnamefont {J.~K.}\ \bibnamefont {Nunes}},\
  and\ \bibinfo {author} {\bibfnamefont {H.~A.}\ \bibnamefont {Stone}},\ }\href
  {https://doi.org/10.1103/PhysRevLett.125.264502} {\bibfield  {journal}
  {\bibinfo  {journal} {Phys. Rev. Lett.}\ }\textbf {\bibinfo {volume} {125}},\
  \bibinfo {pages} {264502} (\bibinfo {year} {2020})}\BibitemShut {NoStop}%
\bibitem [{\citenamefont {True}\ and\ \citenamefont
  {Crimaldi}(2017)}]{true2017}%
  \BibitemOpen
  \bibfield  {author} {\bibinfo {author} {\bibfnamefont {A.~C.}\ \bibnamefont
  {True}}\ and\ \bibinfo {author} {\bibfnamefont {J.~P.}\ \bibnamefont
  {Crimaldi}},\ }\href {https://doi.org/10.1103/PhysRevE.95.053107} {\bibfield
  {journal} {\bibinfo  {journal} {Phys. Rev. E}\ }\textbf {\bibinfo {volume}
  {95}},\ \bibinfo {pages} {053107} (\bibinfo {year} {2017})}\BibitemShut
  {NoStop}%
\bibitem [{\citenamefont {Klamkin}(1971)}]{ellipse02}%
  \BibitemOpen
  \bibfield  {author} {\bibinfo {author} {\bibfnamefont {M.~S.}\ \bibnamefont
  {Klamkin}},\ }\href {http://www.jstor.org/stable/2317530} {\bibfield
  {journal} {\bibinfo  {journal} {The American Mathematical Monthly}\ }\textbf
  {\bibinfo {volume} {78}},\ \bibinfo {pages} {280} (\bibinfo {year}
  {1971})}\BibitemShut {NoStop}%
\bibitem [{\citenamefont {White}(2016)}]{white2016}%
  \BibitemOpen
  \bibfield  {author} {\bibinfo {author} {\bibfnamefont {F.~M.}\ \bibnamefont
  {White}},\ }\href@noop {} {\emph {\bibinfo {title} {Fluid Mechanics, 8th
  Edition}}}\ (\bibinfo  {publisher} {McGraw-Hill Education,},\ \bibinfo {year}
  {2016})\BibitemShut {NoStop}%
\bibitem [{\citenamefont {Tang}\ \emph {et~al.}(2017)\citenamefont {Tang},
  \citenamefont {Li}, \citenamefont {Dong}, \citenamefont {Wang}, \citenamefont
  {Mei},\ and\ \citenamefont {Hu}}]{smoke_TANG2017}%
  \BibitemOpen
  \bibfield  {author} {\bibinfo {author} {\bibfnamefont {F.}~\bibnamefont
  {Tang}}, \bibinfo {author} {\bibfnamefont {L.}~\bibnamefont {Li}}, \bibinfo
  {author} {\bibfnamefont {M.}~\bibnamefont {Dong}}, \bibinfo {author}
  {\bibfnamefont {Q.}~\bibnamefont {Wang}}, \bibinfo {author} {\bibfnamefont
  {F.}~\bibnamefont {Mei}},\ and\ \bibinfo {author} {\bibfnamefont
  {L.}~\bibnamefont {Hu}},\ }\href
  {https://doi.org/https://doi.org/10.1016/j.applthermaleng.2016.08.224}
  {\bibfield  {journal} {\bibinfo  {journal} {Applied Thermal Engineering}\
  }\textbf {\bibinfo {volume} {110}},\ \bibinfo {pages} {1021 } (\bibinfo
  {year} {2017})}\BibitemShut {NoStop}%
\end{thebibliography}%


\providecommand{\noopsort}[1]{}\providecommand{\singleletter}[1]{#1}%

\end{document}